\documentclass[journal=jacsat,manuscript=article]{achemso}

\usepackage[version=3]{mhchem} 

\author{Mirko Rocci}
\email{mirko.rocci@sns.it}
\author{Giorgio De Simoni}
\email{giorgio.desimoni@sns.it}
\author{Claudio Puglia}
\altaffiliation{Department of Physics ”E. Fermi”, Università di Pisa, Largo Pontecorvo 3, I-56127 Pisa, Italy}
\author{Davide Degli Esposti}
\altaffiliation{Department of Physics ”E. Fermi”, Università di Pisa, Largo Pontecorvo 3, I-56127 Pisa, Italy}
\author{Elia Strambini}
\author{Valentina Zannier}
\author{Lucia Sorba}
\author{Francesco Giazotto}
\email{francesco.giazotto@sns.it}
\affiliation[CNRNANO]{NEST, Istituto Nanoscienze-CNR and Scuola Normale Superiore, I-56127 Pisa, Italy}

\title{Gate-Controlled Suspended Titanium Nanobridge Supercurrent Transistor}

\begin{document}


\begin{abstract}
  Under standard conditions, the electrostatic field-effect is negligible in conventional metals and was expected to be completely ineffective also in superconducting metals. This common believe was recently put under question by a family of experiments, which displayed full gate-voltage-induced suppression of critical current in superconducting all-metallic gated nanotransistors. To date, the microscopic origin of this phenomenon is under debate and trivial explanations based on heating effects given by the negligible electron leakage from the gates should be excluded. Here, we demonstrate the control of the supercurrent in fully suspended superconducting nanobridges. Our advanced nanofabrication methods allows to build suspended superconducting Ti-based supercurrent transistors which show ambipolar and monotonic full suppression of the critical current for gate voltages of $V_G^C\simeq18$\,V and  for temperatures up to $\sim$80$\%$  of the critical temperature. The suspended device architecture minimizes the electron-phonon interaction between the superconducting nanobridge and the substrate, and therefore it rules out any possible contribution stemming from charge injection into the insulating substrate. Besides, our finite element method simulations of vacuum electron tunneling from the gate to the bridge and thermal considerations rule out the cold-electron field emission as possible driving mechanism for the observed phenomenology. Our findings promise a better understanding of the field effect in superconducting metals.
  
\end{abstract}


Field effect transistors (FETs) represent one of the fundamental building blocks of the modern electronics era. The electrostatic gating effect is at the basis of FET operation, and is exploited in countless applications. It relies on the strong modulation of the carrier density of the transistor conduction channel by means of an electric field, which is established through the application of a control voltage to a gate electrode. The fairly low carrier density of semiconductor materials is, therefore, pivotal for the realization of any conventional FET-based device, since it prevents the electric field to be unavoidably canceled at the channel surface due to the electrostatic screening. On the contrary and for the same reason, field effect is low or negligible in metals due to their large carrier density, and it was believed to be almost ineffective on superconducting metals as well\cite{Glover1960,Bonfiglioli1962}. With these premises, to overcome the limitations imposed by heat dissipation to semiconductor-based electronics, the search for non-dissipative field-effect devices was, till recently, focused mostly on non-metallic low charge-density superconducting materials, such as high-critical-temperature \cite{Nishino1989,Fiory1990,Mannhart1993b,Okamoto1992a,Mannhart1993_B} and proximity-based \cite{Clark1980,Takayanagi1985,Akazaki1995} superconductors. In these systems, the superconducting order parameter can be modulated as the result of a gate-induced tuning of the charge density. Therefore, the modifications of several superconducting properties, such as the critical temperature and the critical current, are the consequence of a conventionally-intended field effect, and are unipolar, \textit{i. e.} odd in the gate voltage.

Recently, this landscape has been completely repainted by a family of experiments \cite{DeSimoni2018,PaolucciRev2019} carried on all-metallic supercurrent  nano-transistors showing a surprising unconventional gating effect, which cannot be explained by the usual voltage-driven charge density modulation. These include the full ambipolar suppression of the critical supercurrent \cite{DeSimoni2018,Paolucci2018UltraEfficientSD,paolucci2019magnetotransport,bours2020unveiling,desimoni2019josephson,Puglia2020Van}, the increase of quasiparticle population \cite{Alegria2020}, the manipulation of the superconducting phase \cite{Paolucci2019Interf}, and the broadening of the switching current distributions \cite{Puglia2020}. Aside from the high potential for future applications~\cite{PaolucciRev2019}, these findings raised fundamental questions on the origin of these phenomena~\cite{virtanen2019superconducting}, which were not predicted by the Bardeen-Cooper-Schrieffer (BCS) and  Fermi-liquid theories\cite{Larkin1963}. To date, two complementary hypotheses are under debate: an electrostatically-triggered orbital polarization at the superconductor surface~\cite{mercaldo2019,bours2020unveiling}, and the injection of highly-energetic quasiparticles extracted from the gate~\cite{Ritter2020,Alegria2020}.
Here, we tackle this crucial issue \textit{via} a fully suspended gate-controlled Ti nano-transistor. Our geometry allows us to eliminate any direct injection of quasiparticles through the substrate, thereby making cold electron field emission through the vacuum the only possible charge transport mechanism. With the aid of a fully numerical 3D model, based on a finite elements method, in combination with the observed phenomenology, and through thermal considerations we can rule out also the occurrence of cold electron field emission. Excluding these two trivial phenomena is pivotal in light of understanding the microscopic nature of the gating effect in superconducting metallic nanostructures, which represents an unsolved puzzle in contemporary superconductivity. Yet, from the technological point of view, our suspended fabrication technique provides the enabling technology to implement a variety of applications and fundamental studies combining, for instance, superconductivity with nano-mechanics \cite{mcdermott2018strong}.

\begin{figure}[ht]
  \includegraphics[width=0.6\columnwidth]{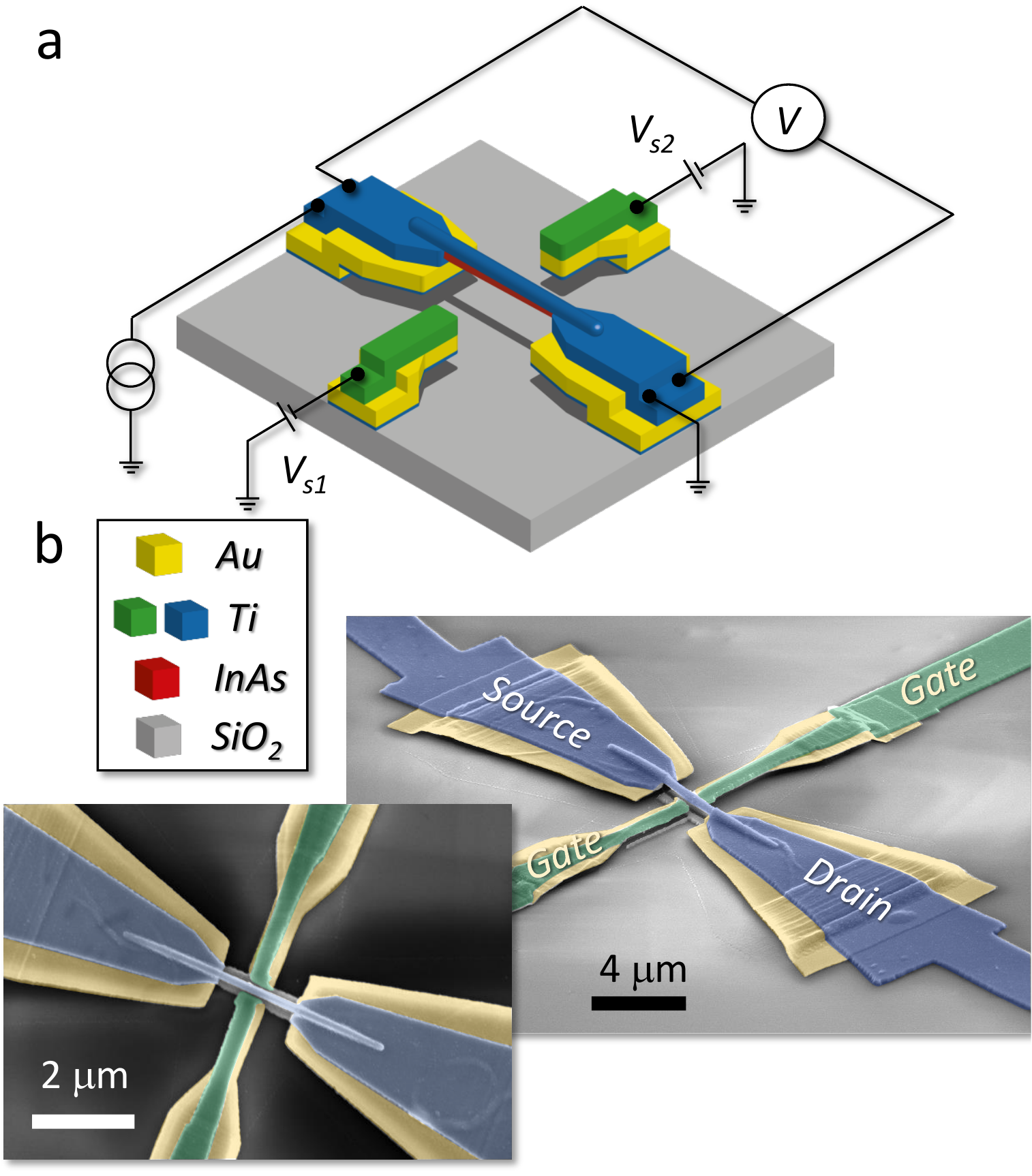}
    \caption{\textbf{Geometry of a suspended Ti gated superconductor transistor.} \textbf{a} 3D sketch showing a suspended Ti-based nanoscale GST. The Ti wire (blue) is deposited on top of suspended InAs nanowire (red) and it is measured in a conventional four-wire configuration. The amplitude of the GST critical supercurrent was controlled by applying a voltage $V_G\equiv V_{s1}=V_{s2}$ to the two side-gate electrodes (green). \textbf{b} False-color scanning electron micrographs (top view and 35$^{\circ}$ tilted, left and right respectively) of a typical device, laid on an intrinsic SiO\textsubscript{2} substrate (grey). The Ti wire is 1.7\,$\mu$m long, 70\,nm thick, $\sim$120\,nm wide, and $\sim$200\,nm raised from the substrate. The two gate electrodes are at $\sim 40$\,nm with respect to the Ti suspended bridge. The gold pads (yellow) are used to hold the suspended structures while the Ti nanowire is mechanically supported by an InAs nanowire located underneath the Ti wire. The InAs nanowire does not contribute to the conduction.}
  \label{fig:fig1}
\end{figure}

\begin{figure}
  \includegraphics[width=\textwidth]{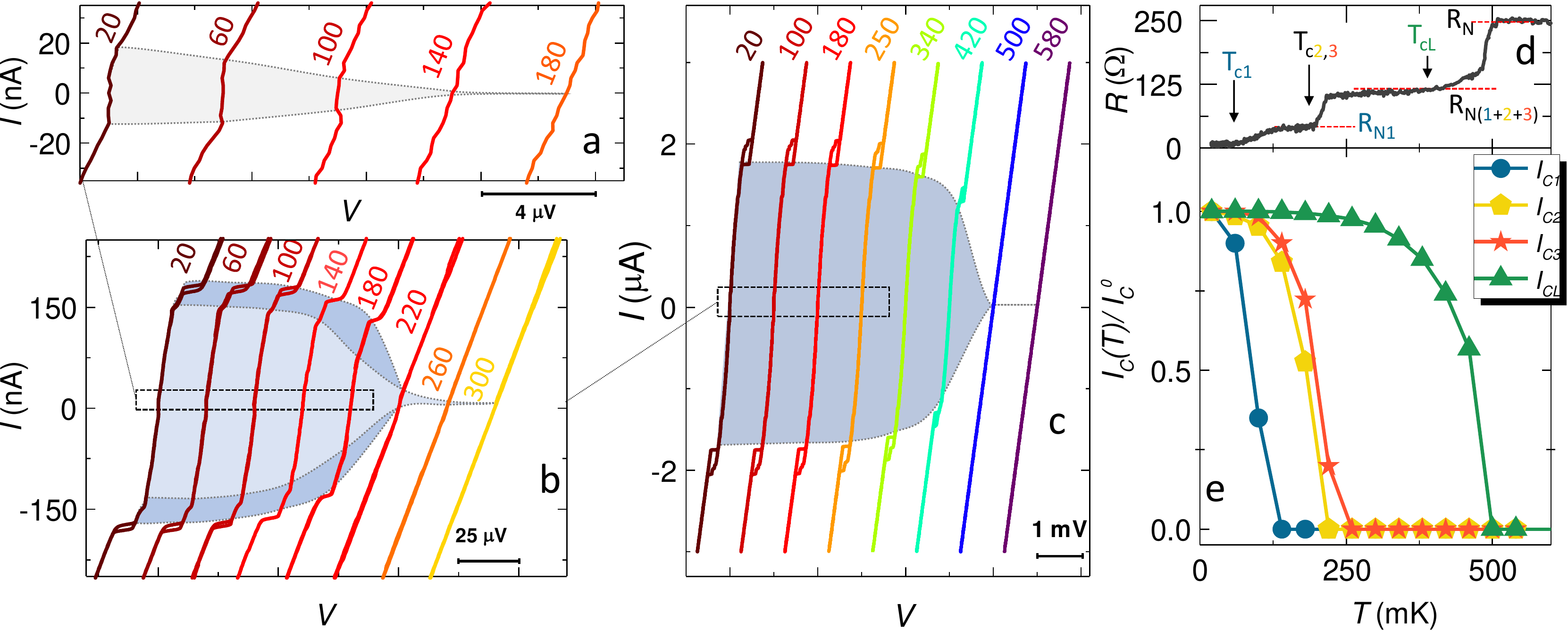}
    \caption{\textbf{Critical currents and critical temperatures of the suspended Ti GST.} 
    \textbf{a} Current-voltage ($I-V$) forward and backward characteristics of the device at several bath temperatures ranging from 20 to 180\,mK. The curves are horizontally shifted for clarity. Dissipationless transport and the evolution of the critical current $I_{c1}$ are highlighted by the light gray area.
    \textbf{b} Blow-out of $I-V$ shown in \textbf{a}, corresponding to the area enclosed by the dotted-line rectangle. Two additional superconducting transitions of the weak links with critical current $I_{c2}$ and $I_{c3}$ are visible and highlighted by the gray and dark-gray areas, respectively. 
    The superconducting behavior disappears for temperatures $T \geq 220$\,mK. 
    \textbf{c} Full range of the $I-V$ at several bath temperatures ranging from 20 to 580\,mK. A fourth hysteretic transition is visible at large currents ($I_{CL} \simeq 2\mu$A), and is consistent with the transition of the Ti leads visible up to $\sim420$\,mK. The curves are horizontally shifted for clarity.
    \textbf{d} Resistance \textit{vs.} temperature ($R-T$) characteristic measured with a lock-in amplifier with a small bias current of $I=1.5$\,nA. 
    The $R-T$ curve displays two sharp and one broad resistance drops ($T\simeq500$\,mK, $T\simeq220$,mK and $T\simeq150$\,mK, respectively) corresponding to the four superconducting critical temperatures of the whole device: $T_{c1}\simeq140$\,mK and $T_{c2}\simeq T_{c3}=220$\,mK which belong to the suspended Ti GST, and $T_{cL}$ which corresponds to the drain-source leads of the suspended Ti wire. These temperatures were determined by using the $R_N/2$ criterion, where $R_N=R_{N1(2)}$ is the resistance value $R$ taken at the plateau within the range of $R(T=T_{c1(2)})$ to $R(T=T_{c2(3)})$. 
    \textbf{e.} Temperature evolution of the critical currents. For a comparison, the curves have been normalized [$I_C(T)/I_C^0$($T=20$\,mK)].}
  \label{fig:fig2}
\end{figure}

\section{Results and discussion}
The geometry of a typical suspended Ti-based gated superconductor transistor (GST) is depicted in Fig.\ref{fig:fig1}a. The devices consist of a 70-nm-thick and $1.7$-$\mu$m-long single suspended Ti nanobridge flanked by two side gate electrodes (green in Fig.\ref{fig:fig1}), separated from the Ti bridge by a gap of $\sim 40$\,nm. All the measurements presented in the following were carried out on the same representative device, where two superconducting source-drain leads connected to the bridge were used to perform low-noise four-terminal transport characterizations in a filtered He$^3-$He$^4$ dilution refrigerator, as schematically depicted in Fig.\ref{fig:fig1}. Our GSTs rely on an \textit{ad-hoc} nanofabrication process conceived to ensure 
a mechanically robust suspension.
The latter was achieved \textit{via} 
an undoped crystalline InAs nanowire \cite{Iorio2019}
set on two pillars of cross-linked insulating PMMA employed as a support scaffold for the fragile Ti layer e-beam evaporated above (blue-colored in Fig. \ref{fig:fig1}). A thermally evaporated Ti/Au (thicknesses 5 and 15 nm, respectively) bi-layer is used to anchor the nanowire to the PMMA and to the substrate (in yellow in Fig. \ref{fig:fig1}). The native oxide of the InAs nanowire and its negligible residual charge guarantee the electrical insulation between the nanowire and the Ti layer. 
The width of the GSTs is not lithographically determined, but depends on the diameter of the InAs nanowire, usually between  80 and 130\,nm. A detailed description of the device fabrication procedure is reported in the Methods section.

At 20 mK, the current-voltage ($I-V$) characteristic of the GST exhibits dissipation-less Cooper pair transport, with a critical supercurrent $I_{C1}\simeq 25$ nA (see Fig. \ref{fig:fig2}a). At higher current, three additional transitions  can be identified,  which stem from the consequential 
transition of two further regions of the bridge, having higher critical currents (at 20 mK, $I_{C2}=150$\,nA,\, $I_{C3}=180$\,nA), and the transition of the superconducting leads ($I_{CL}\simeq 1.8\,\mu$A at 20 mK)
The three different transitions of the nanobridge can be ascribed to inhomogeneities of the Ti film surface, grown on top of a InAs nanowire, and likely generating a series of three weak links along the bridge. 
Figure \ref{fig:fig2}a, b and c show the $I-V$ characteristics at selected temperatures ranging from 20 to 580 mK.
Each $I_C(T)$ monotonically decays \cite{Bardeen1967} (see Fig. \ref{fig:fig2}e), and vanishes in correspondence of the respective critical temperature ($T_c$). 
$T_c$ was also measured by a standard 4-wire lock-in technique of the bridge resistance $R$ \textit{versus}  bath temperature $T$ (see  Fig. \ref{fig:fig2}d). 
Clear steps at $T_{c1}\simeq150$\,mK, $T_{c2}\simeq T_{c3}\simeq280$\,mK and $T_{cL}\simeq 500$\,mK were observed, and correspond to the transition of the three weak links and of the leads, respectively. Above the Ti film transition temperature $T_{cL}$, the GST shows a normal state resistance of $R_N\simeq250$\,$\Omega$. 

\begin{figure}
  \includegraphics[width=0.6\columnwidth]{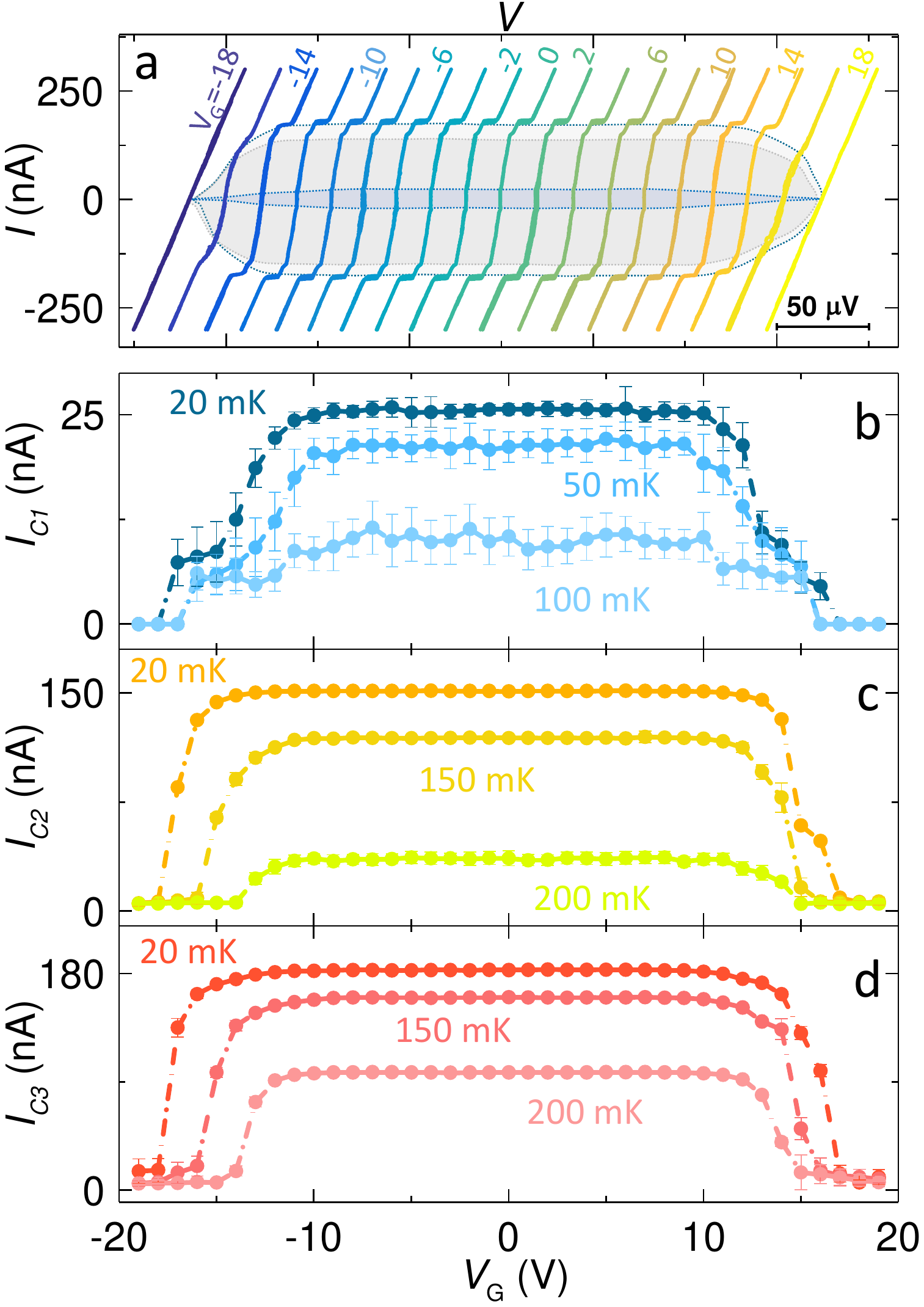}
  \caption{\textbf{Response of the suspended Ti-based nano-GST to the gate voltage.}  \textbf{a} Current\textit{vs.} voltage characteristics (back and forth) for  different  applied gate voltage ($V_G$) values measured at $T_{Bath}=20$\,mK. The curves are horizontally offset for clarity. 
  Blue and grey areas, and the dotted grey lines are guides to the eye highlighting the gate-driven evolution of $I_{C1}$, $I_{C2}$ and $I_{C3}$.  \textbf{b} Switching current $I_{C1}$ \textit{vs.} $V_G$ measured at selected bath temperatures $T_{Bath}$. Similarly to (b), \textbf{c} and \textbf{d} show the $V_G$ dependence of the critical current $I_{C2}$ and $I_{C3}$, respectively. The error bars represent the standard deviation of $I_{C1,2,3}$ calculated over 25 repetitions.}
    \label{fig:fig3}
\end{figure}

In analogy to gating experiments on non-suspended devices, we observed gate effects on $I_C$ by measuring the evolution of the $I-V$ characteristics at different voltages ($V_G$) applied to both side gate electrodes (see Fig.\ref{fig:fig1}a).
Such symmetric configuration minimizes the mechanical strain generated by the Coulomb interaction between the bridge and the gates, which may damage or destroy the whole device. 
Figure \ref{fig:fig3}a displays the $I-V$ characteristics measured at 20\,mK for selected $V_G$ values. 
For $V_G$ exceeding $\pm$12\,V all the critical currents of the weak links show a monotonic suppression, down to the full quenching, for both positive and negative voltage values, thereby confirming the superconducting gate effect also in suspended structures. Moreover, the weak-link normal-state resistance is totally unaffected by gating. We emphasize that the gate voltage independently and directly affects all the three weak links. Indeed, when $I_{c1}<I<I_{c2},I_{c3}$ and no gate voltage is applied, the measured device resistance is equal to $R_1$, corresponding to a situation in which the second and the third weak links persist in the superconducting state, while the first is resistive. By applying and then increasing the gate voltage at constant current bias, weak links 2 and 3 are forced to switch to the normal state [\textit{i.e.} $I_{c2}(V_G),I_{c3}(V_G)<I$] and a device resistance equal to $R_{1+2+3}$ is measured. This consideration let us to exclude that the transition to the normal state of the weak links with critical current $I_{c2}$ and $I_{c3}$ was caused by an overheating driven by the transition to the normal state of the weak link with the lowest critical current. Therefore, although the microscopic origin of such phenomenology is still under debate, we guess the gate voltage to be responsible for a direct action on the switching current probability distribution of each weak link, through the induction of phase slips in the whole suspended wire region\cite{Puglia2020}.

From the $I-V$ curves we extracted the $I_C-V_G$ characteristics for the three weak links, which are displayed in Fig. \ref{fig:fig3}b, c and d for selected bath temperatures. All of them exhibit the usual dependence on $V_G$, where a plateau at low gate voltages is followed by a sudden drop of the critical current above the gate voltage threshold $\vert V_G\vert\simeq12$\,V. 
Full quench of the supercurrent of all the weak links was observed for $V_G\sim \pm 18$ V. As the temperature increases, the plateau amplitude lowers, stemming from the decay of the critical current with temperature, and the plateau width shrinks, reaching at 200 mK a width of about 80$\%$ of the one at 20 mK. 
For $I_{C2}$ and $I_{C3}$ this effect is accompanied by a reduction of the critical current pinch-off voltage of the same percentage (see Figs. \ref{fig:fig3}c and d). The latter observations are in contrast with earlier works \cite{DeSimoni2018,Paolucci2018UltraEfficientSD,paolucci2019magnetotransport,PaolucciRev2019}, where the plateaus widened and the pinch-off voltages were constant by increasing the temperature. 
We ascribe this difference to the reduced bridge-to-substrate thermal coupling with respect to devices laying on a substrate.
Indeed, independently of the microscopic origin of the gate effect, the suppression of the critical supercurrent seems always to be associated to a substantial enhancement of the number of quasi-particles present in the superconducting wire \cite{Alegria2020,Puglia2020}. This increase is likely to be more effective in suspended transistors where relaxation of quasi-particle excitations \textit{via} electron-phonon interaction is reduced with respect to non-suspended devices.

\begin{figure*}{t}
  \includegraphics[width=0.8\textwidth]{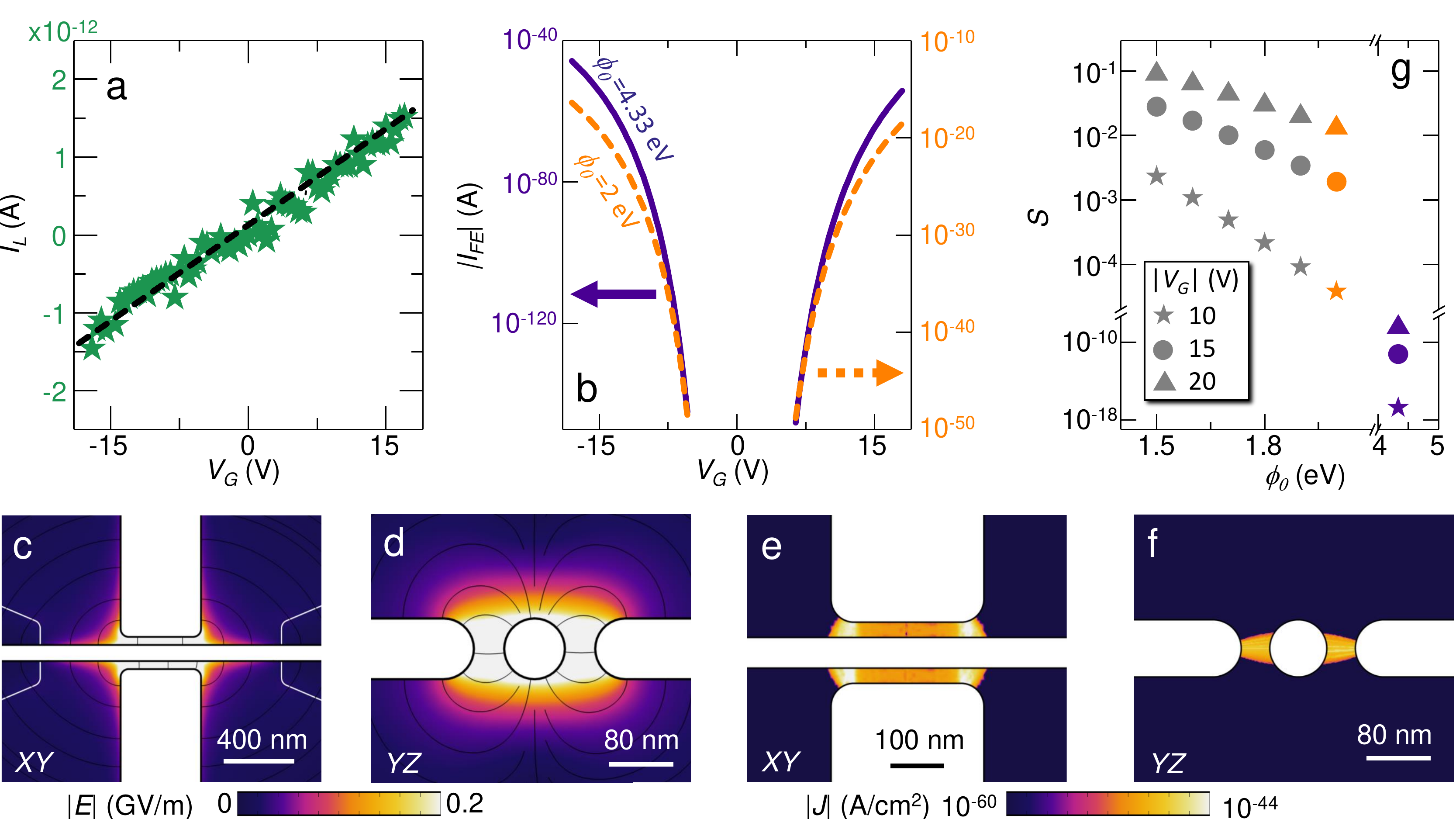}
    \caption{
    \textbf{Electric field spatial distribution and cold field emission current.}
    \textbf{a} Leakage current ($I_L$) between the two side gates and the suspended Ti nano-GST, measured at 20 mK as a function of $V_G$.
    \textbf{b}: Current-voltage characteristics $I_{FE}-V_G$ obtained \textit{via} numerical integration  of the Fowler-Nordheim current density ($J_{cathode}$, Eq.~\ref{eqFN}) at the cathode surface for $\phi_0=4.33$ eV (Ti, violet curve, left axis) and $\phi_0=2$ eV (orange curve, right axis).  
    \textbf{c,d} Electric field module ($\vert \textbf{E}(x,y,z)\vert$) and streamline on an \textit{XY} and \textit{YZ} cut-planes. $E_{x,y,z}$ is calculated for an applied voltage of -15 V. The cut-planes are centered with respect to the suspended nanobridge. The distribution reveals the concentration of the field-effect on the bridge. 
    \textbf{e,f} Current density module $\vert \textbf{J}(x,y,z)\vert$ evaluated on a \textit{XY} and \textit{YZ} cut-planes, obtained by solving the electron ballistic transport from the gates surfaces toward the Ti bridge (and vice versa for opposite $V_G$). Here we set $V_G=-15$ V, and $\phi_0=4.33$ eV. 
    The current distributions show how the injection mechanism barely affects a 500\,nm portion of the nano-bridge. The bridge and gates volumes (white) in \textbf{c}, \textbf{d}, \textbf{e} and \textbf{f} are modeled as perfect conducting domains. 
    \textbf{g} Log-scale plot of the symmetry factor $S(|V_G|)=|I_{FE}(|V_G|)|/|I_{FE}(-|V_G|)|$ calculated for $|V_G|$ equal to 10 V (stars), 15 V (dots), and 20 V (triangles). Points corresponding to $\phi_0=4.33$ eV (Ti) and $\phi_0=2$ eV (\textit{i. e.}, calculated at the same values of the curves in panel \textbf{b}) are colored in violet and orange, respectively.}
  \label{fig:fig4}
\end{figure*}

To quantify the small leakage current flowing between the gates and the bridge ($I_L$) a high-gain room-temperature current pre-amplifier was connected to the bridge referred to ground. 
The $I_L-V_G$ characteristic is linear (see Fig. \ref{fig:fig4}a), corresponding to a conductance of $\sim10^{-13}$ \,$\Omega^{-1}$, and the current is lower than $\sim 1.5$ pA in the full range of the applied voltage. We stress that such a value is comparable with those obtained in previous experiments performed on non-suspended devices  set on sapphire substrates \cite{DeSimoni2018,bours2020unveiling,Puglia2020,de2020niobium}, thereby suggesting that most of the measured 
leakage current is likely to be dispersed through the wiring setup. 

Differing from previous works, the peculiar suspended architecture of our devices allows to make some precise assessments about the spatial distribution of $I_L$: in the first place, $I_L$ can  propagate through the substrate only from the gate to the side leads. This consideration allows to exclude any local Joule overheating transferred to the bridge \textit{via} phonon coupling caused by a leakage current injected through the substrate. In the second place, a direct flow of current from the gate to the bridge (and vice versa) would be possible only \textit{via} cold-electron field-emission (CFE) through vacuum.  
The latter might be expected to occur due to the application of an intense electric field between the gate electrodes and the bridge.
To shed light on the role played by an eventual field-emitted electronic current in the $I_C$\ quenching, we numerically quantified the CFE current ($I_{FE}$) by means of 3-dimensional finite-element method simulations performed on a geometry equivalent to our real device, and compared it with the measured $I_L$. A detailed description of the whole simulation procedure is reported in the Methods section. 

$I_{FE}$
was calculated by integrating on
the cathode surface, \textit{i. e.}, the gate(wire) for negative(positive) $V_G$, the Fowler-Nordheim tunnel current density that at the cathode reads\cite{fowler1928electron,simmons1963generalized} 
\begin{equation}
\label{eqFN}
J_{cathode}\left[E(x,y,z), \phi_{0}\right]=\frac{2.2 e^{3} E^{2}}{8 \pi h \phi_{0}} \exp \left[-\frac{8 \pi}{2.96 h e E}\left(2 m_{e}\right)^{1 / 2} \phi_{0}^{3 / 2}\right],
\end{equation}
    where \textit{E(x,y,z)} is the amplitude of the electric field on the cathode surface, \textit{m}\textsubscript{e} is the electron mass, $e$ is the electron charge, \textit{h} is the Plank's constant, and $\phi_0= 4.33$\,eV is the work function of  Ti \cite{polyTiworkfunction}. The electric field vector $\mathbf{E}(x,y,z;V_G)$ was previously calculated in the three-dimensional vacuum space region, surrounding the side gates and the bridge, through the Maxwell equation $\mathbf{E}=-\nabla V(x,y,z;V_G)$, where the potential $V(x,y,z;V_G)$ was obtained 
    from the numerical integration of the Poisson equation $\nabla^2 V(x,y,z;V_G)=0$. The bridge and the gate electrode surfaces were simulated with perfect equipotential conductor boundaries set at $V=0$ and $V=V_G$, respectively. 
The spatial distribution of the electric field module $\vert$\textbf{\textit{E}}$\vert$ obtained for the whole 3-dimensional domain of the simulation is color plotted in Fig. \ref{fig:fig4}c,d in top (\textit{XY}-plane) and cross-section (\textit{XZ}-plane) views for $V_G=-15$\,V. 
The electric field is localized between the Ti nanobridge and the side gate surfaces while it quickly vanishes elsewhere, therefore not affecting the leads. 
Combining the information on $\mathbf{E}$ and of
$J_{cathode}$ allows to calculate the full spatial distribution of the current density [$\textbf{\textit{J}}(x,y,z)$] by resolving the ballistic trajectories of the electrons emitted by the cathode.
The color plots in Fig. \ref{fig:fig4}e,f show the top and cross-section views of $|\textbf{\textit{J}}(x,y,z)|$, calculated for $\phi_0=4.33$\,eV and $V_G=-15$ V. The resulting particle trajectory plots indicate a highly-localized electron emission on the Ti nanobridge surface, in correspondence of the gate-bridge gap. Thus, in the case of  occurrence of CFE, the totality of the electrons emitted from the gate (or from the bridge) is absorbed from the bridge (or from the gate).

By integrating the current density over the cathode surface yields $I_{FE}(V_G)$, as displayed in Fig. \ref{fig:fig4}b (violet curve, right scale) along with the plot of $I_L$. 
Notably, $I_{FE}$ is many orders of magnitude smaller than the maximum gate-bridge leakage current experimentally measured, which is most likely injected through the substrate into the leads, or dispersed toward the insulation of the wiring. 
This suggests that an eventual CFE current should be not measurable within the electric field scales of our experiment. According to our calculations, this current would correspond 
to the emission of one electron every $10^{28}$ years on average,
and is consistent with an electric field at the surface of the cathode which is too weak to trigger a proper CFE current.
Indeed, cold-electron emission generally requires $E$ at least of the order of $1\div 10$ GV/m \cite{Joag2012}, while in our case the maximum electric field at the cathode surface is
one order of magnitude smaller, at most. 

The incompatibility between the field emission hypothesis and our experimental observations is further supported by the substantial asymmetry of $\vert I_{FE} \vert -V_G$ in contrast to the  symmetry of $I_C-V_G$. Indeed, due to the exponential dependence of CFE on $E$ (see Eq. \ref{eqFN}) joined with the non-symmetric geometry of the cathode electrodes of GST devices~\cite{DeSimoni2018,PaolucciRev2019}, a symmetric $\vert I_{FE} \vert -V_G$ is not plausible. This is made evident by our simulations where  $I_{FE}$ is suppressed by several orders of magnitude for positive $V_G$ values with respect to negative ones. 
 Even assuming a substantial underestimate of the CFE mechanism in our model, the latter consideration remains valid despite any arbitrary choice of the model parameters like, e. g., the work function (see the orange curve in Fig. \ref{fig:fig4}b, right scale, calculated for $\phi_0=2\,$eV). 
This issue can be quantified by defining a symmetry factor $S(V_G)=|I_{FE}(|V_G|)|/|I_{FE}(-|V_G|)|$, shown in Fig. \ref{fig:fig4}g \textit{vs.} work function $\phi_0$ for selected gate voltage values. These plots allow to appreciate how, in the gate-voltage range in which $I_C$ suppression occurs, $I_{FE}$ remains non symmetric, with the $S$-parameter reaching at most  $\sim0.1$ for a non-realistic work function $\phi_0=1.5$ eV. This latter analysis makes therefore very unlikely a direct relation between an eventual CFE current and the gate-voltage ambipolar suppression of $I_C$, which was universally observed in the present and previous experiments performed on all-metallic supercurrent transistors.

Finally, the inconsistency of the CFE hypothesis comes as well from simple thermodynamic arguments. 
Indeed, as discussed in detail in references \cite{Puglia2020,Giazotto2006a,Timofeev2009}, 
the emission of a ballistic electron from the gate to the bridge should release into the superconducting wire an energy of several eVs.
Such a process results in a sudden increase of the electronic temperature quantified by the relation $T_f=\sqrt{\frac{2eV_G}{\Omega \gamma}+T_{i}^2}$, where $\Omega$ the volume of the bridge, $\gamma$ the Sommerfeld's constant of Ti, and $T_f$ and $T_i$ are the final and initial electronic temperatures, respectively. For the absorption of a single electron emitted at $V_G=5$ V, (\textit{i. e.}, a value for which no $I_C$ suppression was ever observed) a sudden increase of $T_f\sim600$ mK is expected for $T_i=20$ mK, which is  $\sim 20\%$ higher than $T_{cL}$. It therefore follows that a single highly-energetic electron absorption would result in a sudden destruction of the superconducting state, which is incompatible with the smooth damping of $I_C-V_G$~\cite{Puglia2020}. Furthermore, for positive gate voltage, electrons are field-emitted from the bridge around the Fermi level, and their energy is released into the gates. The thermodynamics of electron emission and absorption, therefore, is very different, and the two processes occur at energy scales extremely uneven so that it turns out difficult to reconcile them with the observed bipolar $I_C$ suppression with gate voltage.  

\section{Conclusions}
In conclusion, our cutting-edge suspended device architecture allowed us to take a different perspective compared to previous studies, and to investigate the effect of applied electrostatic fields on the superconducting properties of a  nano-GST. 
Our experiments allow to unequivocally exclude any current injected through the insulating substrate as a possible trigger of the GST.
Moreover, our analysis demonstrated that cold-electron field-emission between the gates and the bridge is very unlikely to occur, and does not play any obvious role in the physical description of the  supercurrent suppression process. 
These evidences remark that the still elusive fundamental microscopic mechanisms at the basis of the phenomenon have to be addressed. Yet, the generality of our fabrication protocol provides a technological platform enabling the investigation of a variety of groundbreaking suspended all-metallic-based GSTs with applications in superconducting nanoelectronics and spintronics \cite{linder2015superconducting}. The latter may also benefit by the creation of future paradigms and device concepts, such as exchange-coupled triplet paired GSTs and gate-tunable superconducting spin-filter Josephson junctions based on EuS/Al and NbN/GdN multilayered heterostructures \cite{deSimoni2018EuS,Senapati2011spinJ-NbN-GdN,diesch2018creation,pal2017oddfreqNbNGdN}, as well as gate controlled topological superconductivity \cite{vaitiekenas_zero-field_2020}.


\section*{Methods}
\subsection{Fabrication Process}

\begin{figure}
  \includegraphics[width=\textwidth]{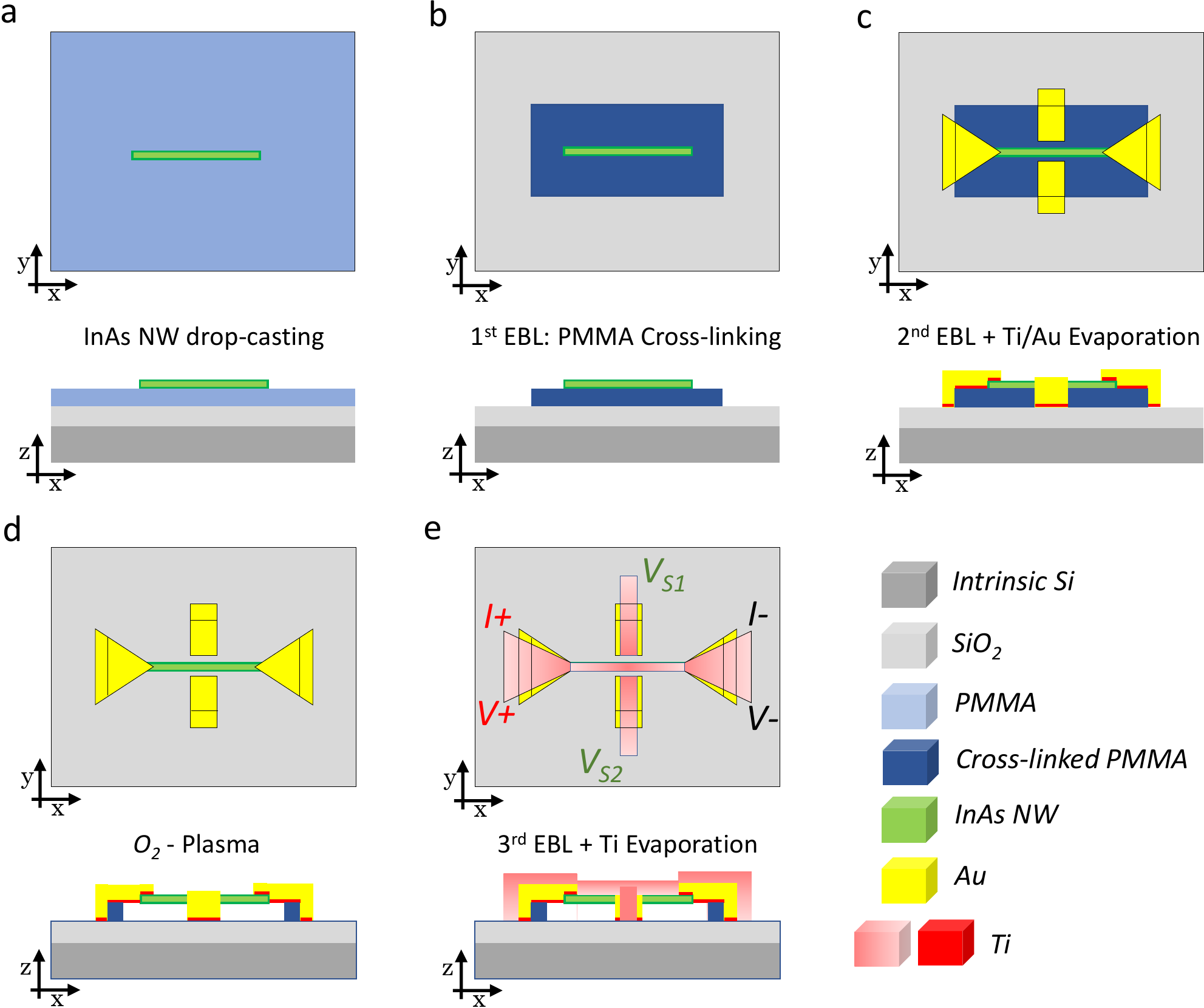}
  \caption{\textbf{Suspended Ti-nanoFETs Nanofabrication process steps.} \textbf{a.} InAs NWs dropcasting onto 200\,nm thick PMMA. \textbf{b.} EBL step for the PMMA cross-linking around the targeted InAs NW. \textbf{c.} EBL and thermal evaporation of Ti/Au (5/15\,nm) for the pattern definition of the drain-source and gate supporting areas. \textbf{d.} $O_2-$Plasma for the cross-linked PMMA removal. \textbf{e.} EBL and electron-beam evaporation of 70\,nm$-$thick Ti nanobridge, side-gates and leads.}
  \label{fig:fig1s}
\end{figure}

The Ti GST fabrication involved the use of undoped InAs self-assembled crystalline nanowires (NWs) grown by chemical beam epitaxy (CBE)\cite{gomes2015controlling}. A five-step fabrication process was developed to achieve the Ti bridge suspension. First, the NWs were drop-casted onto a 200\,nm-thick PMMA layer (AR-P 679.04 from Allresist GmbH) covering a SiO\textsubscript{2}/intrinsic-Si substrate (see Fig. \ref{fig:fig1s}a). Afterwards, a high-dose electron beam lithography (EBL) exposure (5000 $\mu$C/cm$^2$ at 10 keV) was carried out to cross-link the PMMA\cite{zailer1996crosslinked} underneath the NW. The cross-linking process of the polymer chains makes the over-exposed PMMA insoluble to any conventional PMMA remover ($i.e.$, acetone). 
The InAs NW suspension was then accomplished by immersing the sample in acetone for 10\,min and rinsing it in IPA. This step allows us to remove the unexposed PMMA, keeping intact the cross-linked one (see Fig. \ref{fig:fig1s}b). Our sample was then subjected to a second re-aligned EBL exposure, followed by a thermal evaporation of a Ti/Au (5/15\,nm) bi-layer. The resulting drain and gate areas are visible in Fig. \ref{fig:fig1s}c (in yellow). Such a process results in an efficient anchoring of the InAs NW to the cross-linked PMMA and the substrate by a continuous Au film. Reactive O\textsubscript{2}$-$plasma etching (for 10\,min at 100\,W) was then applied to remove the uncovered cross-linked PMMA portion\cite{teh2003cross}. Due to the chemical inertness of both the Au pads and the InAs NW to the reactive $O_2$ etching process, only the unprotected cross-linked PMMA was removed (Fig. \ref{fig:fig1s}d). This gives rise to a large undercut along the Au pad edges\cite{teh2003cross,Iorio2019}, which is crucial to prevent any undesired short circuit among the electrodes, after the Ti deposition. The GST fabrication was finalized with an EBL nanopatterning of the Ti nanobridge, side gates and micrometric leads, followed by an electron beam evaporation of 70\,nm-thick Ti layer, performed at room temperature in an ultra-high vacuum evaporator (base pressure: $\sim10^{-11}$\,Torr;  deposition rate: 10$-$13\,\r{A}/s). The final device structure is visible in Fig. \ref{fig:fig1s}e.

\subsection{Finite element simulations}

The computational results are obtained via a four-step finite-element model (FEM) simulation, where the geometrical parameters were set up consistently with the typical suspended Ti-based nanodevice dimensions (see Fig. 1 of the main text). The gate-bridge distance, along the $y-$axis was set equal to 40\,nm. The nanobridge was schematized as a $1.7-\mu$m-long hollow cylinder parallel to the $x$-axis and with a radius of 80 nm. 
The simulation domain coincides with the vacuum region surrounding the device, and consisted in a box with $x-y-z$ sides of $3\times 2 \times 0.5$ $\mu$m$^3$. The wire and gate surfaces constituted a hollow equipotential boundary within the box. A tetrahedral mesh was used to discretize the domain volume, with a minimum and a maximum distance between the nodes of 0.25\,nm and 100 nm, respectively.

In the first simulation step the potential $V(x,y,z;v_G)$ was obtained numerically integrating the Poisson equation $\nabla^2 V(x,y,z;V_G)=0$ over the entire simulation domain for selected values of the gate voltage parameter $V_G$, with the potential boundary condition of $V=V_G$ and $V=0$ on the side gates and bridge surfaces, respectively.  The electric field distribution $\mathbf{E}(x,y,z;V_G)$ was then calculated (second step) through the Maxwell equation $\mathbf{E}=-\nabla V(x,y,z;V_G)$. The electric field on the gates or the bridge surface, depending on the sign of $V_G$, was substituted into the Fowler-Nordheim equation to calculate the CFE current density, and its integration  over the cathode surface leads to the total emitted current. Finally (fourth step), the surface current density was exploited to solve the equation of motion of the electrons traveling between the gates and the nanobridge.

In order to ensure the reliability of the results, a mesh convergence study was performed, using  the maximum value of the electric field modulus on the electrodes surfaces and the total emitted current as checkpoints.


\begin{acknowledgement}

The authors acknowledge the Horizon 2020 innovation programme under Grant Agreement No. 800923-SUPERTED, and Marie Skłodowska-Curie grant agreement EuSuper No. 796603. The authors thank  M. Aprili, A. Braggio, M. Cuoco, S. Gasparinetti, F. Paolucci, and P. Solinas for fruitful discussions.

\end{acknowledgement}

\section*{Author Contributions}
M.R. and C.P. fabricated the devices, and performed the experiment 
with input from G.D.S., E.S., and F.G.. V.Z. and L.S. grew the InAs nanowires.
D.D.E., M.R., and C.P. analyzed the data with input from all the authors.  D.D.E. performed the simulations with inputs from G.D.S., E.S., and F.G.. M.R. and G.D.S. wrote the manuscript with input from all the authors. F.G. conceived the experiment. All of the authors discussed the results and their implications equally.



\providecommand{\latin}[1]{#1}
\makeatletter
\providecommand{\doi}
  {\begingroup\let\do\@makeother\dospecials
  \catcode`\{=1 \catcode`\}=2 \doi@aux}
\providecommand{\doi@aux}[1]{\endgroup\texttt{#1}}
\makeatother
\providecommand*\mcitethebibliography{\thebibliography}
\csname @ifundefined\endcsname{endmcitethebibliography}
  {\let\endmcitethebibliography\endthebibliography}{}

\end{document}